\documentclass[aps,prmaterials,twocolumn,10pt,showpacs,preprintnumbers,amsmath,amssymb,superscriptaddress,longbibliography]{revtex4-2}

\usepackage{epsfig}                                                            
\usepackage{graphicx}
\usepackage{dcolumn}
\usepackage{color}
\usepackage{bm}
\usepackage{subcaption}
\usepackage[colorlinks=true,
linkcolor=blue,
citecolor=blue,
urlcolor=blue]{hyperref}
\usepackage{fnpct}

\newcommand{\be}{\begin{equation}} \newcommand{\ee}{\end{equation}}
\newcommand{\bea}{\begin{eqnarray}} \newcommand{\eea}{\end{eqnarray}} 
\begin{document}
	
	\title{\bf Origin of bias-voltage-dependent transient negative capacitance in PZT ferroelectric capacitor with labyrinthine-like domains}
	
	\author {Ganga S. Kumar$^{\dagger}$ }\email{Corresponding author}  \affiliation {Multiscale Microstructure and Mechanics of Materials (4M) Division, CSIR-Central Glass and Ceramic Research Institute, Kolkata 700032, India}
	\author {Kusumpal Yadav} \thanks{These authors have contributed equally} \affiliation {School of Physical Sciences, Indian Association for the Cultivation of Science, Kolkata 700032, India}
	\author {Miral Verma} \thanks{These authors have contributed equally} \affiliation {School of Minerals, Metallurgical and Materials Engineering, Indian Institute of Technology, Bhubaneswar 752050, India}
	\author {Parushottam Majhi} \affiliation {Department of Physics, Indian Institute of Technology, Bombay, Mumbai 400076, India}
	\author {Pragati Sharma} \affiliation {UGC-DAE Consortium for Scientific Research, Indore 452001, India}
	\author {Kaustuv Chatterjee} \affiliation {Multiscale Microstructure and Mechanics of Materials (4M) Division, CSIR-Central Glass and Ceramic Research Institute, Kolkata 700032, India}
	\author {Sandip Bysakh} \affiliation {Multiscale Microstructure and Mechanics of Materials (4M) Division, CSIR-Central Glass and Ceramic Research Institute, Kolkata 700032, India}
	\author {Sudipta Goswami} \affiliation {School of Materials Science and Nanotechnology, Jadavpur University, Kolkata 700032, India}
	\author {Prabir Pal} \affiliation {Multiscale Microstructure and Mechanics of Materials (4M) Division, CSIR-Central Glass and Ceramic Research Institute, Kolkata 700032, India}
	\author {Chandan K. Ghosh} \affiliation {School of Materials Science and Nanotechnology, Jadavpur University, Kolkata 700032, India}
	\author {S. B. Roy} \affiliation {Ramakrishna Mission Vivekananda Eductaional and Research Institute, Howrah 711202, India}
	\author {Vasant Sathe} \affiliation {UGC-DAE Consortium for Scientific Research, Indore 452001, India}
	\author {Devajyoti Mukherjee} \affiliation {School of Physical Sciences, Indian Association for the Cultivation of Science, Kolkata 700032, India}
	\author {Dipten Bhattacharya} \affiliation {Multiscale Microstructure and Mechanics of Materials (4M) Division, CSIR-Central Glass and Ceramic Research Institute, Kolkata 700032, India}
	
	\date{\today}
	
	\begin{abstract}
		We investigated how the bias-voltage-dependent variation in the ferroelectric domain switching pathways affects the transient negative capacitance in multidomain PZT ferroelectric capacitor with labyrinthine-like domains having, primarily, 90$^\circ$ walls. Using the vector piezoresponse force microscopy, x-ray photoelectron spectroscopy, and energy dispersive x-ray spectroscopy, we show that, together with strain, local variation in composition results in the formation of labyrinthine domains in epitaxial films of thickness $\sim$1 $\mu$m. Clear signature of emergence of bias-voltage-dependent pathways could be observed in the Raman modes exhibiting splitting due to an intermediate state - different from the initial and final states - at the ``intermediate state of switching". The polarization relaxation patterns also show different states emerging depending on the bias voltage characteristics.  The phase-field simulation of relaxation of ``intermediate state of switching" corroborates the experimental results and also exhibits dependence of the domain-wall area on the bias voltage and damping parameter. Combining the results of experiments and theoretical calculation, we show, therefore, that the variation of bias voltage leads to a change in the domain switching pathways and each pathway is associated with specific domain-wall area at the ``intermediate state of switching". Since, the net domain-wall area at the ``intermediate state of switching" governs the magnitude of transient negative capacitance, this work thus traces the origin of tunability of transient negative capacitance in PZT ferroelectric capacitor.  
	\end{abstract}
	
	\maketitle
	
	\section{Introduction}
	
	For more than ten years now extensive research has been carried out globally on identifying and stabilizing the negative capacitive state of a ferroelectric capacitor, offering a pathway towards energy-efficient electronic devices
	\cite{iniguez2019ferroelectric,hoffmann2019why}. The concept of NC has been explored in two regimes -- the quasi-steady static negative capacitance (SNC)\cite{zubko2016negative,saha2021negative,yadav2019spatially}, stabilized under appropriate boundary conditions in ferroelectric/dielectric heterostructures, and the transient negative capacitance (TNC), which is a non-equilibrium polarization state arising due to the dynamic competition between nucleation dynamics and the lattice driven structural transitions, modifying the free energy landscape during polarization \cite{khan2015negative,hoffmann2019unveiling,li2024dual}. Scores of novel applications such as negative-capacitive-field-effect transistor (NC-FET)\cite{salahuddin2008use}, electrostatic supercapacitor\cite{hoffmann2019negative,cheema-2024}, electromechanical actutators \cite{Raman,graf2025elastic} etc., have been identified during this period which effectively exploits the benefits of NC. Beyond conventional device scaling, recently, NC has been proposed as a tool for Coulomb engineering \cite{shankar2025exploiting}, by tuning of electron-electron interactions. The theoretical predictions of NC-mediated superconductivity \cite{datta2021can}, hints to a possibility that ferroelectric systems could provide a novel route to stabilize superconducting ground states in low-dimensional materials thereby opening new avenues of application potential. 
	
	In a multidomain system, variation of polarization and polarizability across the domain-domain-wall region yields negative capacitance at the local scale because of locally opposing polarization  \cite{lukyanchuk2019harnessing,zubko2016negative}. Therefore, domain-wall area turns out to be governing the magnitude of negative capacitance in such systems  \cite{hoffmann2018domain,park2019modeling}. Theoretical simulation of the role of thickness of the films, types of domains, domain switching mechanism - homogeneous versus inhomogeneous, domain switching speed and the corresponding viscosity \cite{xu2020enhanced,cao2024multiple,eliseev2022observability} etc., has earlier been carried out. However, a fundamental issue - whether the magnitude of the negative capacitance can be tuned by varying the bias voltage amplitude and frequency and, if so, why the negative capacitance exhibits such tunability - has not been investigated in detail so far.  
	
	In this paper, we demonstrate that TNC is tunable and is directly governed by the bias-voltage-dependent domain switching pathways. We have determined the types of domains experimentally and show from Raman spectra recorded under a carefully designed bias voltage protocol - which helps in collecting the Raman spectroscopy at the positive and negative saturated polarized states as well as at the switching states around the coercive voltage - that the domain switching pathways vary depending on the bias voltage. The relaxation of the intermediate state of switching has also been tracked and found to be dependent on the bias voltage. Phase-field simulation of the relaxation not only corroborates the experimental results but also determines how the net domain-wall area versus time patterns - i.e., the relaxation patterns - depend on the bias voltage. Combining the experimental and the theoretical results, we show that variation in the bias voltage parameters results in ferroelectric domain switching following different pathways and each pathway is associated with specific domain-wall area and switching speed. Since, the domain-wall area during switching governs the magnitude of the TNC, it turns out to be tunable via variation of the bias voltage. This insight could be useful in generating and freezing a ``switching" state to yield very large and tunable SNC as well. 

	\begin{figure*}[t]
		\begin{center}
		{\includegraphics[ width=\textwidth]{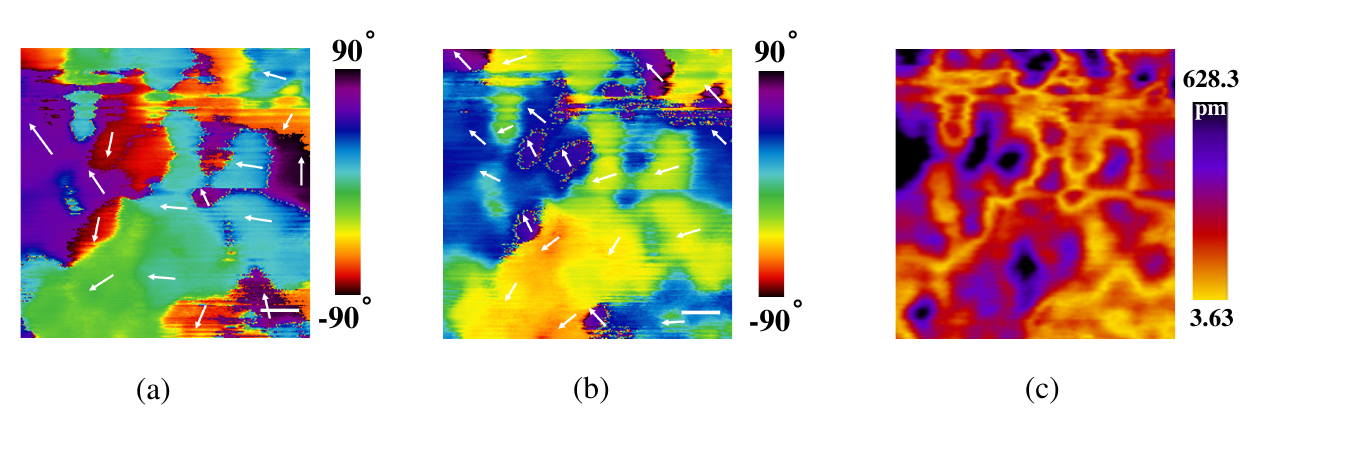}}
                   \end{center}
		\caption{PFM images showing (a) lateral phase, (b) vertical phase, and (c) amplitude images of (111) oriented PZT film. The arrows in the image shows the locally inferred direction of polarization from the phase contrast image. Scale bar : 200 nm. }
	\end{figure*}

\begin{figure}[b]
		\begin{center}
		{\includegraphics[width=0.5\textwidth]{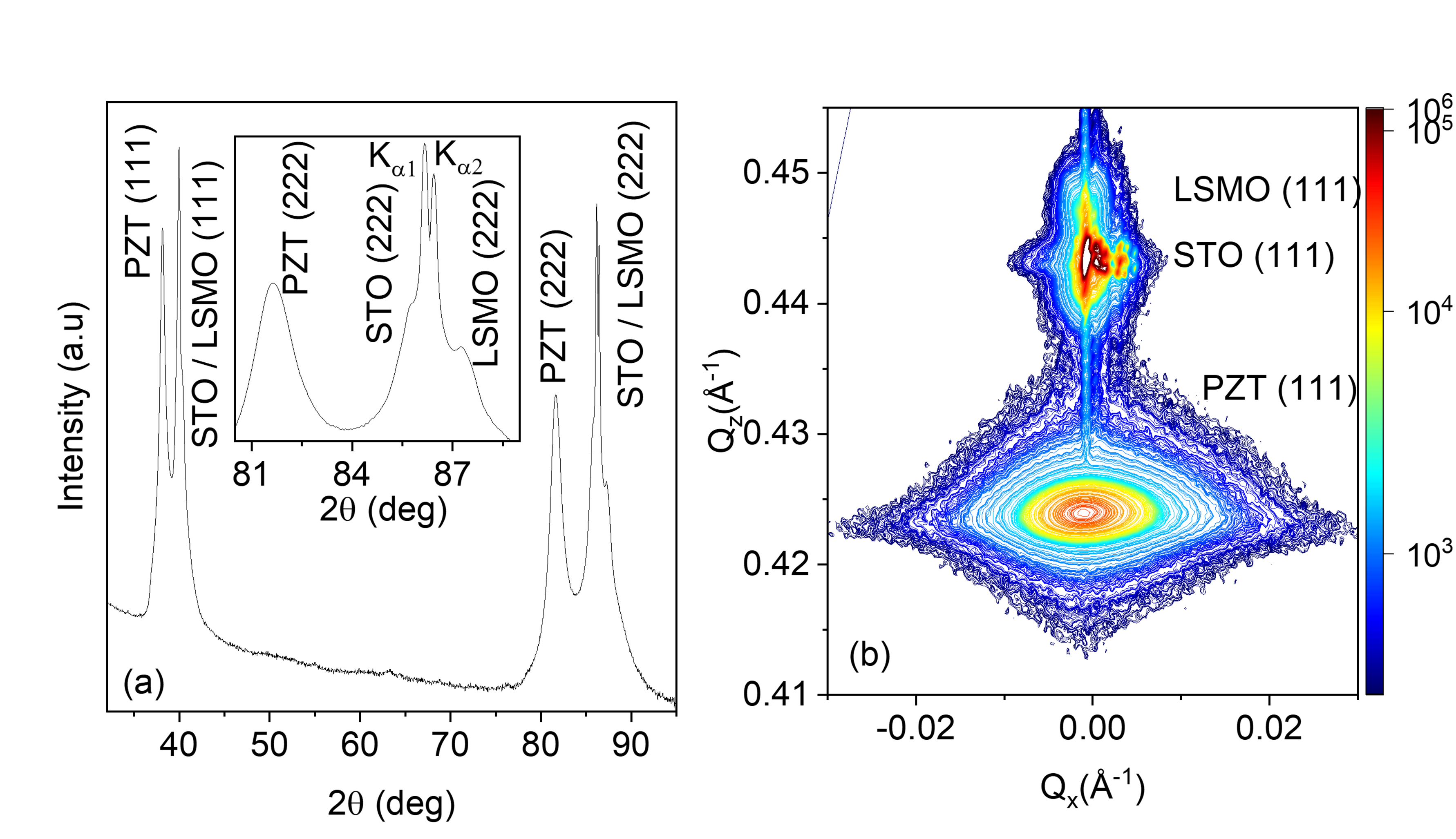}}
                  \end{center}
		\caption{(a) X-ray diffraction ($\theta$–2$\theta$) pattern of the PZT/LSMO/STO (111) heterostructure, showing the diffraction reflections assigned to the constituent layers and STO substrate. (b) Reciprocal space map (RSM) acquired around the STO (111) reflection to investigate the structural characteristics of the heterostructure.}
	\end{figure}

	\section{Experimental Details}
	
	Epitaxial film of Pb(Zr,Ti)O$_3$ with half-metallic La$_{0.7}$Sr$_{0.3}$MnO$_3$ (LSMO) as the top and bottom electrodes was grown on (111) oriented single crystal SrTiO$_3$(STO) substrates by Pulsed Laser Deposition (PLD). The thickness of film as determined from the transmission electron microscopy (TEM) images of the cross-sectional sandwiched thin films is $\sim$1 $\mu$m. The crystallinity of the thin-film heterostructures was characterized using a Rigaku Smart Lab 9 kW X-ray diffractometer equipped with a five-axis goniometer sample stage and using collimated, parallel-beam Cu $K\alpha$ radiation ($\lambda$ = 1.5406 $\AA$). The film was, as well, characterized by using  x-ray photoelectron spectroscopy (XPS) line scan and energy dispersive x-ray spectroscopy (EDS) in conjunction with field-emission scanning electron microscope (FESEM) and transmission electron microscope (TEM) to observe the local variation in the composition. The local crystallographic structure has been determined by using selected area diffraction pattern (SAED) technique in transmission electron microscope. Vector piezoresponse force microscopy (VPFM) has been used to image the ferroelectric domains. XPS was recorded using the PHI 5000 Versa Probe-II x-ray spectrophotometer. A focused X-ray source of Al-K$\alpha$ (1486.7 eV) was used to probe the core levels, in an ultra-high vacuum environment of a 3.7 $\times$ 10$^{-8}$ Pa. The calibration for carbon peak to ensure the charge correction has been done at 284.7 eV. The vector piezoresponse force microscopy images were recorded using  a commercial AFM from Asylum Research with Pt coated conductive probe (resonance frequency $\sim$ 70 kHz and spring contstant $\sim$ 2 Nm$^{-1}$) at a scan speed of 1 Hz to obtain simultaneous out-of-plane and in-plane piezoresponse signals. Raman spectroscopy under a bias voltage was carried out to identify if there is emergence of an intermediate phase during switching.The spectra was recorded in the range 100-900 cm$^{-1}$ using Horiba Jobin-Yvon HR 800 micro-Raman spectrometer in backscattering geometry with an 1800 g/mm grating. A 473 nm laser (3 mW) was focused through a microscope objective to a spot of  $\sim$ 2 $\mu$m. Influence of bias-voltage-dependent domain switching pathways was tracked by studying the relaxation of polarization corresponding to the ``intermediate state of switching". The ferroelectric  polarization (P)--voltage (V) hysteresis loops were measured by employing bipolar triangular pulses in the ferroelectric hysteresis loop tracer from Radiant Inc. (Precision LC II). The polarization relaxation has been investigated by using a precisely designed  positive-up-negative-down (PUND) protocol. All the electrical pulse measurements were carried out by using probe station equipped with tungsten probes (Radiant Inc.).
	
	\begin{figure*}[t]
		\centering
		{\includegraphics[width=0.95\textwidth]{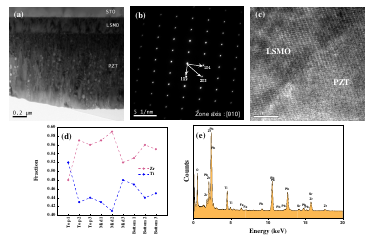}}
		\caption{(a) cross-sectional TEM micrograph showing the PZT/LSMO/STO layers, (b) SAED pattern of the PZT layer recorded along [010] zone axis, (c) HRTEM image of the PZT/LSMO interface, (d) Zr-Ti fraction calculated from the EDS obtained from different points across the PZT layer, (e) Representative EDS spectrum of the PZT layer showing the different elements present. }
	\end{figure*}
	
	\section{Computational Details}
	
	The evolution of the ferroelectric domain structure and its response to the applied bias voltage were investigated using a two-dimensional time-dependent Ginzburg–Landau (TDGL) phase-field model. The simulations were performed on a square computational domain by solving for the spatially and temporally varying polarization field, [P(x,y,t)]. The model was formulated to capture the formation and evolution of ferroelectric domains and their response to a time-dependent electric field. The total free-energy density of the ferroelectric system was expressed as
	
	\begin{equation}
		u = aP^2 + bP^4 + cP^6 - EP + k(\nabla P)^2
	\end{equation}
	
	where a, b, and c are Landau expansion coefficients, E is the applied electric field, and k is the gradient energy coefficient related to polarization inhomogeneity.
	
	The first three terms describe the local ferroelectric free-energy landscape, whereas the gradient term governs the spatial variation of polarization and determines the characteristic domain-wall width. The electric-field coupling term accounts for the interaction between the polarization and the applied bias.
	
	The temporal evolution of the polarization field was obtained by solving the Landau–Khalatnikov equation,
	
	\begin{equation}
		\frac{\partial P}{\partial t} = -\frac{1}{\rho}\frac{\delta F}{\delta P}
	\end{equation}
	
	where $\rho$ is the damping parameter which represents internal loss and F is the total free energy of the system.
	
	The effective electric field E$_\text{eff}$ was computed as
	
	\begin{equation}
		E_{\text{eff}} = \frac{V_{\mathrm{F}}}{t_{\mathrm{F}}} - E_{\text{bias}}
	\end{equation}
	
	where V$_{\mathrm{F}}$ is the voltage across the ferroelectric, t$_{\mathrm{F}}$ is the film thickness, and E$_{\mathrm{bias}}$ represents internal bias fields due to interfacial asymmetries at the electrodes.
	
	\begin{figure*}[t]
		\centering
		{\includegraphics[width=0.7\textwidth]{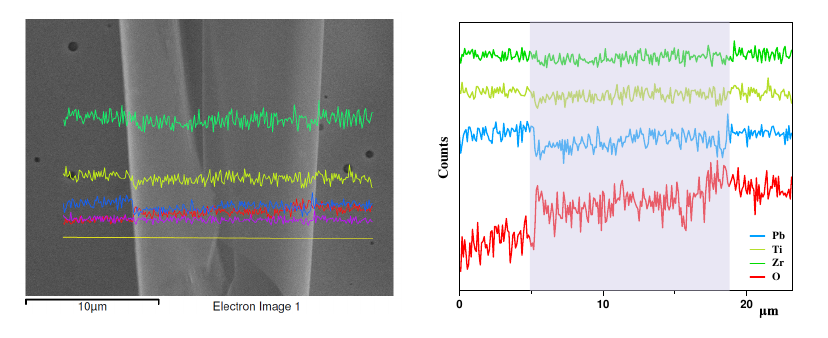}}
		\caption{The FESEM image and EDS line scan profile of the PZT film with the highlighted region showing compositional fluctuation.}
	\end{figure*}
	
	\section{Results and Discussion}
	
	\subsection{Types of domains}
	
	The vector piezoresponse force microscopy (PFM) with simultaneous acquisition of vertical and one component of lateral response has been used to identify the types of domains. The corresponding phase-contrast and amplitude images are shown in the Figs. 1(a)-(c). The domains show labyrinthine (maze-like) like structure. The amplitude image of (111) film (Fig. 1(c)) shows a complex domain structure and the amplitude sharply decreases at the domain walls due to the suppression of piezoresponse at the domain walls. The phase image shows different color contrast corresponding to the various directions of the domain orientation. The angle between the boundary of two adjacent domains in the phase images has been measured\cite{sahoo2020improved} for insights about domain wall angle. From the phase-contrast images of (111) film in Fig. 1(a) and (b), we can see that both in-plane and out-of-plane components are present. This suggests that the multiple polarization variants are possible, but a full reconstruction would require acquisition of both lateral components. It can be seen that the majority of the domain walls have an angle 90$^o$ or less, which is due to the presence of coexisting local rhombohedral structure. The labyrinthine or bubble type domains are normally observed in ultrathin films due to poor charge screening, high depolarizing field, mechanical boundary conditions like strain, charge doping and competition between electrostatic and elastic energy\cite{zhang2017nanoscale,govinden2023stability,eliseev2018labyrinthine,nahas2020topology}. In our case, the thickness of the epitaxial film is about 1 $\mu$m, which is well above the ultrathin regime. To evaluate the origin of labyrinthine-like domains we have carried out XRD, TEM, FESEM-EDS and XPS. 

Figure 2(a) presents the X-ray diffraction (XRD) $\theta$–2$\theta$ pattern of the PZT/LSMO/STO (111) heterostructure. The observed diffraction peaks are assigned to the (111)-oriented PZT and LSMO layers and the (111) reflection of the single-crystalline STO substrate. The absence of discernible additional diffraction peaks suggests that no secondary crystalline phases are detectable within the sensitivity of the measurement. These results indicate the preferential out-of-plane orientation of the constituent layers along the pseudocubic [111] direction. The out-of-plane lattice parameters are approximately 3.867 $\AA$ for pseudocubic LSMO (JCPDS No. 01-089-4461) and 4.086 $\AA$ for pseudocubic PZT (JCPDS No. 01-070-6602). To further investigate the reciprocal-space distribution of the diffracted intensity, an RSM was acquired around the symmetric STO (111) reflection [Fig. 1(b)]. The RSM provides additional information on the diffraction features associated with the substrate and the epitaxial layers. The out-of-plane lattice parameters estimated from the RSM are reported to be consistent with those obtained from the $\theta$–2$\theta$ measurements. Taken together, the XRD results support the formation of a highly (111)-oriented PZT/LSMO heterostructure on the STO (111) substrate. 
	
Cross-sectional TEM has been used to study the interface of the individual layers of the PZT/LSMO/STO structure. The PZT layer shows nano-columnar growth of the constituent PZT grains with nanoscale porosity distributed throughout the PZT layer (Fig. 3(a)). The interplanar spacing ($d$) measured from the SAED pattern (Fig. 3(b)) was found to be consistent with the rhombohedral R3c phase (JCPDS 01-092-2777), from which the (10$\bar{2}$), (104), and (202) reflections were indexed and the zone axis is identified to be [010]. The angle between (104) and (10$\bar{2}$) is found to be 89.4$^{\circ}$. The HRTEM image of the PZT/LSMO interface (Fig. 3(c)) reveals a sharp and coherent interface without the presence of any amorphous interfacial layer.The Fig. 2(e) shows the EDS spectrum obtained from the PZT layer confirming the presence of O, Pb, Ti, and Zr. The presence of a weak Sr peak in the EDS spectrum could be due to diffusion of the Sr atoms from the LSMO layer while Fe is from the specimen support stainless-steel ring. The quantitative EDS analysis from different locations across the PZT layer shows (Fig. 3(d)) that the Zr fraction, given by $x$ = Zr/(Zr+Ti), varies between 0.52 to 0.59 across the layer, although the variation is not systematic, showing local compositional variation rather than a compositional gradient. This local variation in Zr/Ti ratio, by Vegard's law, is expected to give a corresponding variation in the relaxed lattice parameter of the film. Since the film is not fully epitaxially constrained and considerable strain relaxation can take place, the compositionally distinct regions may not be able to adopt their equilibrium lattice parameters resulting in the compositional variation probably being accommodated elastically as local strain field within the PZT layer.
	
	The FESEM-EDS line scan data across the PZT film is shown in Fig. 4. The EDS spectra showed presence of La and Mn in addition to Pb, Zr, Ti and O, which could be from the LSMO top electrodes. Along the scan area, Ti and Zr intensities shows uniformity indicating homogeneity of the B site cations whereas the Pb and O intensities show localized variation as shown within the highlighted region. This local inhomogeneity in Pb gives rise to A-site vacancies, which together with O vacancies can introduce local variation in strain and in the screening of depolarizing field, resulting in a spatial variation of built-in field \cite{agar2015complex}. 
	
	The XPS line scan was conducted to map the spatial fluctuations of chemical states along a line across the film. The fundamental core level spectra acquired at three distinct locations within the scanned area of the PZT film are illustrated in Fig. 5. At each location, the Pb 4f, Zr 3d, and Ti 2p energy levels each exhibit a single spin–orbit doublet associated with Pb$^{2+}$, Zr$^{4+}$, and Ti$^{4+}$ charge state. The Ti 2p spectra exclusively show a Ti$^{4+}$ doublet, with no evidence of Ti$^{3+}$, indicating that Ti retains its nominal oxidation state. The O 1s spectra were resolved into three components: lattice oxygen (M-O), surface-adsorbed hydroxyl groups (M-OH), and a vacancy-related component that fluctuates between 6-8\% across the examined region, indicating a finite defect population. This defect-associated oxygen population may influence the local domain wall pinning energy. The oxygen defect identified through XPS, along with local compositional discrepancies observed in TEM–EDS and FESEM–EDS, suggests that the PZT film exhibits spatially heterogeneous chemistry that contributes to the stabilization of the labyrinthine-like domain structure seen in the VPFM. Consequently, these findings imply that the combined data from local compositional fluctuations observed via FESEM and TEM EDS, defect-associated oxygen detected through XPS, and the presence of various phases contribute to the labyrinthine-like domain architecture in our scenario, rather than the inadequate screening and depolarizing field that stabilize these domains in ultrathin films.
	
	\begin{figure*}[htbp]
		\centering
		{\includegraphics[width=\textwidth]{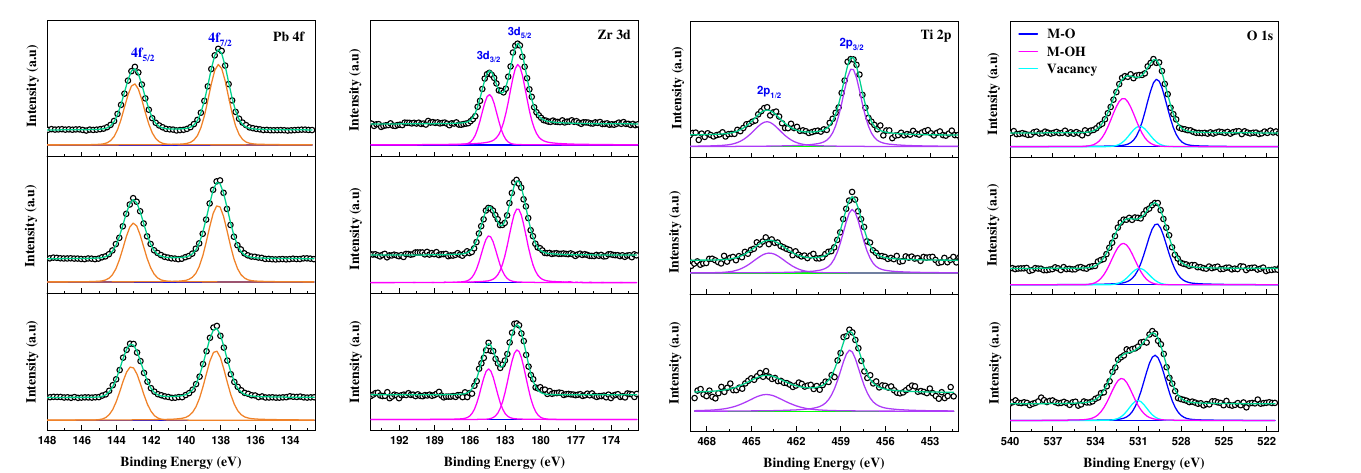}}
		\caption{XPS core level spectra taken at three points along a line for the (111) oriented PZT film.}
	\end{figure*}
	
	\subsection{Consistency of the Negative Capacitance Data}
	
	In order to verify the consistency of the ferroelectric polarization and negative capacitance data, we have measured the P-V hysteresis loops as a function of timescale in the (111) oriented film over many cycles both with or without intermediate heat treatment of the film. Fig. S1 of supplemental materials document\cite{supplementarydoc} shows the major-to-minor hysteresis loops traced for different cycles of measurement. The measurements done on a pristine sample is designated as cycle 1, the second set of measurements without any intermediate heat treatment as cycle 2 and the third set of measurements after applying a heat treatment cycle is designated as cycle 3 from now on. The P-V loops are traced by applying a preset pulse followed by a standard bipolar triangular pulse. The preset pulse ensures that the state of start point in each measurement is identical. 
	
	The minor to major hysteresis loops has been traced for a range of timescale from  0.2 ms to 100 ms and voltage from 2 to 32V in each cycle of measurements. These loops are then differentiated with respect to voltage to get the specific negative capacitance as described in our previous work \cite{kumar2026enhanced}. The specific negative capacitance maximizes for a particular timescale and voltage that is around the coercive voltage and is small for every other timescale and voltage, as in \cite{kumar2026enhanced}, for all the cycles of measurements. The Fig. 6 shows the specific capacitance obtained as a function of time scale for different cycle of measurement. It can be seen that for cycle 1 and cycle 3 the specific negative capacitance maximizes corresponding to timescale 0.5 ms whereas for cycle 2 it maximizes around 5 ms (Fig. S2 of supplemental materials document\cite{supplementarydoc}). This shows that the degradation in the ferroelectric switching properties is probably due to the increase in the defect pinning and charge accumulation as a result of repeated measurement. The properties of the sample can be revived by heating the sample to a temperature above the curie temperature (T$_{\mathrm{C}}$) so as to ensure the redistribution of the defects.
	
	We have also calculated the rate of change of the radius of curvature given by $\approx$ $d$P$^3$/ $d$V$^3$. The rate of curvature change shows how fast the polarization transition takes place around the switching region. It can be seen that the rate of curvature change of the loops undergoes a sharp change from negative to a positive peak value after which it flattens out close to zero showing negligible curvature change. This indicate a rapid transition from unswitched-rapid switching-switched taking place. The negative region corresponds to the unswitched region, the positive peak shows rapid switching where nucleation and growth is rapid and the region of transition from negative to positive represents the regime where switching onsets and nucleation growth is balanced. The near-zero curvature change shows that only fully switched states persist beyond the corresponding time scale. 
	
	The domain switching kinetics has been studied by tracing the change in switched polarization as a function of increasing pulse width for different applied voltage biases employing the Positive-Up-Negative-Down (PUND) pulse method. The plot of normalized switched polarization ($\Delta$P/2P$_{\mathrm{s}}$) against the $log t$ is shown in Fig. 7. The switching behaviour is more appropriately described by the NLS model rather than the KAI model, conforming to our previous observations and reported by others as well \cite{jo2007domain,hu2014universal}. It can be seen that at low voltage (i.e., 5 V), the polarization evolution is very slow and the NLS model shows better fitting indicating nucleation to be rate limiting factor in this voltage regime. For the case of intermediate voltage bias (i.e, 10 V), the switching becomes more rapid compared to V= 5 V, pointing that both nucleation and domain wall motion contribute significantly. At higher voltage (V= 20 V), the NLS model is still valid but the switching kinetics has become close to KAI model, indicating that at voltages higher than this, the switching turns to more homogeneous.
	
	\begin{figure*}[!]
		\centering
		{\includegraphics[width=0.8\textwidth]{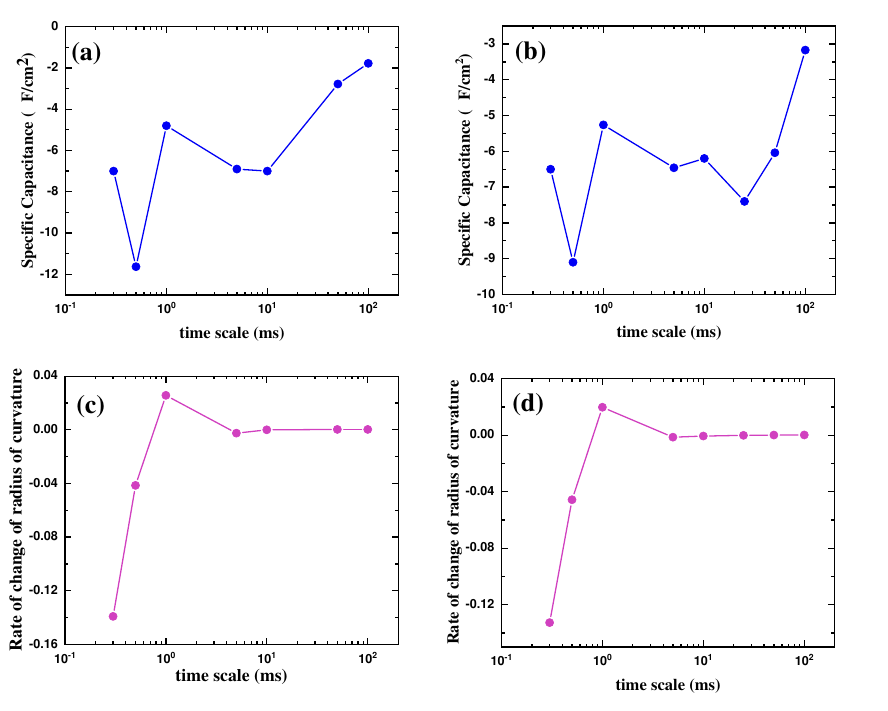}}
		\caption{The specific capacitance plotted as a function of time scale: (a) cycle 1 and (b) cycle 3, and the rate of change of radius of curvature as a function of time scale: (c) cycle 1 and (d) cycle 3.}
	\end{figure*}
	\begin{figure*}[!]
		\centering
		{\includegraphics[width=0.46\textwidth]{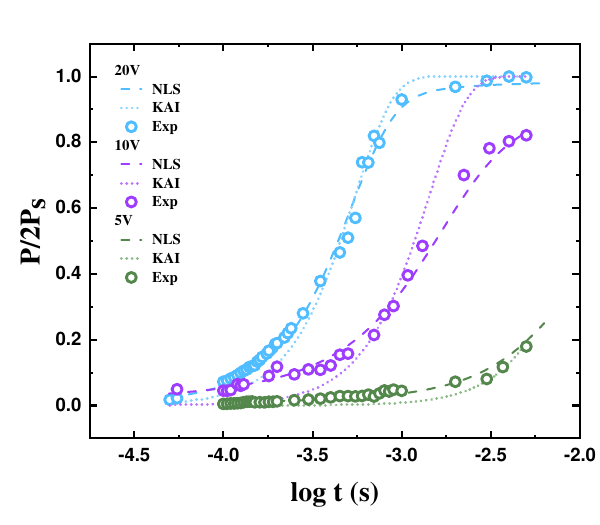}}
		\caption{The evolution of switched polarization as a function of time scale for different applied bias amplitudes. The corresponding fitting uses KAI (dotted lines) and NLS (dashed lines) models.}
	\end{figure*}
	
	\begin{figure*}[!t]
		\centering
		{\includegraphics[width=0.7\textwidth]{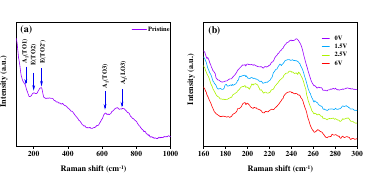}}
		\caption{(a) The Raman spectra corresponding to pristine state and (b) spectra recorded in remnant state after applying bias voltages.}
	\end{figure*}
	It has been shown by the DFT based first principles calculations, carried out  using nudged elastic band (NEB) model, that the ferroelectric domain switching can follow different switching pathways \cite{hu-2025}. Each pathway is associated with an energy barrier and the most favorable pathway of switching is the one with minimum energy barrier. In fact, the domain switching pathway was also shown to be dependent on the bias voltage applied \cite{hu-2025}. We have shown, in our earlier work \cite{kumar2026enhanced}, that in a multidomain sample the variation of bias voltage and time scale can result in different pathways that can give rise to different domain wall length. By tuning the bias voltage parameters to match the intrinsic switching kinetics parameters, we can induce the pathway which would result in maximum domain wall length hence the maximum transient negative capacitance. We have used Raman spectroscopy under electric bias to probe if an intermediate state different from the initial and final state arise at the time of switching in our PZT samples. 
	
	\subsection{Raman spectroscopy under electric bias}
	
	Raman spectra was recorded in prinstine state as well after applying carefully chosen bias voltage. The voltage is applied for 1s and the spectra is recorded, after turning the bias off, in the remanent state. Fig. 8(a) shows the room-temperature Raman spectrum of a (111)-oriented PZT thin film. The Zr fraction has been observed to vary within 0.48 to 0.59 approximately across the film thickness, as determined by nano-scale space-resolved TEM–EDS analysis, the overall composition of the film being within the morphotropic phase boundary (MPB) regime, where rhombohedral and tetragonal ferroelectric phases coexist while and the monoclinic phase acts as an intermediate bridge between them. 
	
	In the pristine (unbiased, 0 V) state, the Raman spectra exhibits predominantly rhombohedral-like characteristics, and the phonon modes are therefore assigned according to rhombohedral PZT symmetry, consistent with the overall Zr-rich composition. For the (111)-oriented film, this geometry allows observation of both A1 and E symmetry Raman modes in accordance with the selection rules.
	
	The low-frequency region (100-220 cm$^{-1}$) is dominated by the (E(TO2)) mode, while the (A(TO1)) mode appears as a weak feature near 160cm$^{-1}$. The (E(TO2)) vibration is widely recognized as a sensitive spectroscopic indicator of the rhombohedral-to-tetragonal structural crossover in PZT near the MPB  \cite{buixaderas2015compositional,Buixaderas2016}. In the pristine state, this mode appears considerably broadened around 200 cm$^{-1}$, reflecting the local structural heterogeneity due to coexistence of multiple ferroelectric phases and compositional fluctuations characteristic of MPB compositions with nearby (A(LO1)) mode further contributing to the spectral broadening in this region \cite{buixaderas2015compositional,frantti-1999}. A broad feature centred near 240 cm$^{-1}$ is assigned to the (E(TO2')) mode which consists of multiple overlapping components \cite{buixaderas2015compositional}. In the high-frequency region ($>$ 550 cm$^{-1}$), the (A1(TO3)) and (A1(LO3)) vibrations merge into a broad spectral envelope, consistent with the Raman response reported for rhombohedral PZT compositions in the vicinity of the MPB \cite{buixaderas2015compositional,deluca-2011}.
	
	The Raman spectra exhibit pronounced changes in the phonon modes within the 180-260 cm$^{-1}$ spectral region under an applied electric field, as shown in Fig. 8(b). The applied voltages were selected to probe the structural evolution at four representative states: 0 V, 1.5 V (below coercive voltage V$_{\mathrm{C}}$), 2.5 V (near V$_{\mathrm{C}}$), and 6 V (well above V$_{\mathrm{C}}$), with V$_{\mathrm{C}}$ determined from the P-V hysteresis loops. At 0 V, the Raman spectrum exhibits a broad and asymmetric feature in the 180-220 cm$^{-1}$ region rather than a single resolved peak, suggesting the coexistence of multiple local structural environments. The broad feature near 240 cm$^{-1}$, attributed to the E(TO2') vibration, also consists of overlapping components\cite{buixaderas2015compositional}. This broadening is linked to the compositional gradient across the thickness, which creates locally distinct vibrational environments through continuous variation of the Zr/Ti ratio.
	
	Upon increasing the applied voltage to 1.5 V, the broad low-frequency feature begins to split into distinguishable components, accompanying a symmetry-lowering structural evolution. Such spectral behaviour is consistent with an increasing contribution from a monoclinic phase or an electric-field-induced redistribution of the phase fractions toward a monoclinic-like state. As the applied voltage approaches the coercive voltage (V$_{\mathrm{C}}$ $\sim$ 2.5 V), the relative intensity of the higher-wavenumber component gains relative intensity, suggesting enhanced stabilization or growth of monoclinic domains.
	\begin{figure*}
		\centering
		{\includegraphics[width=0.45\textwidth]{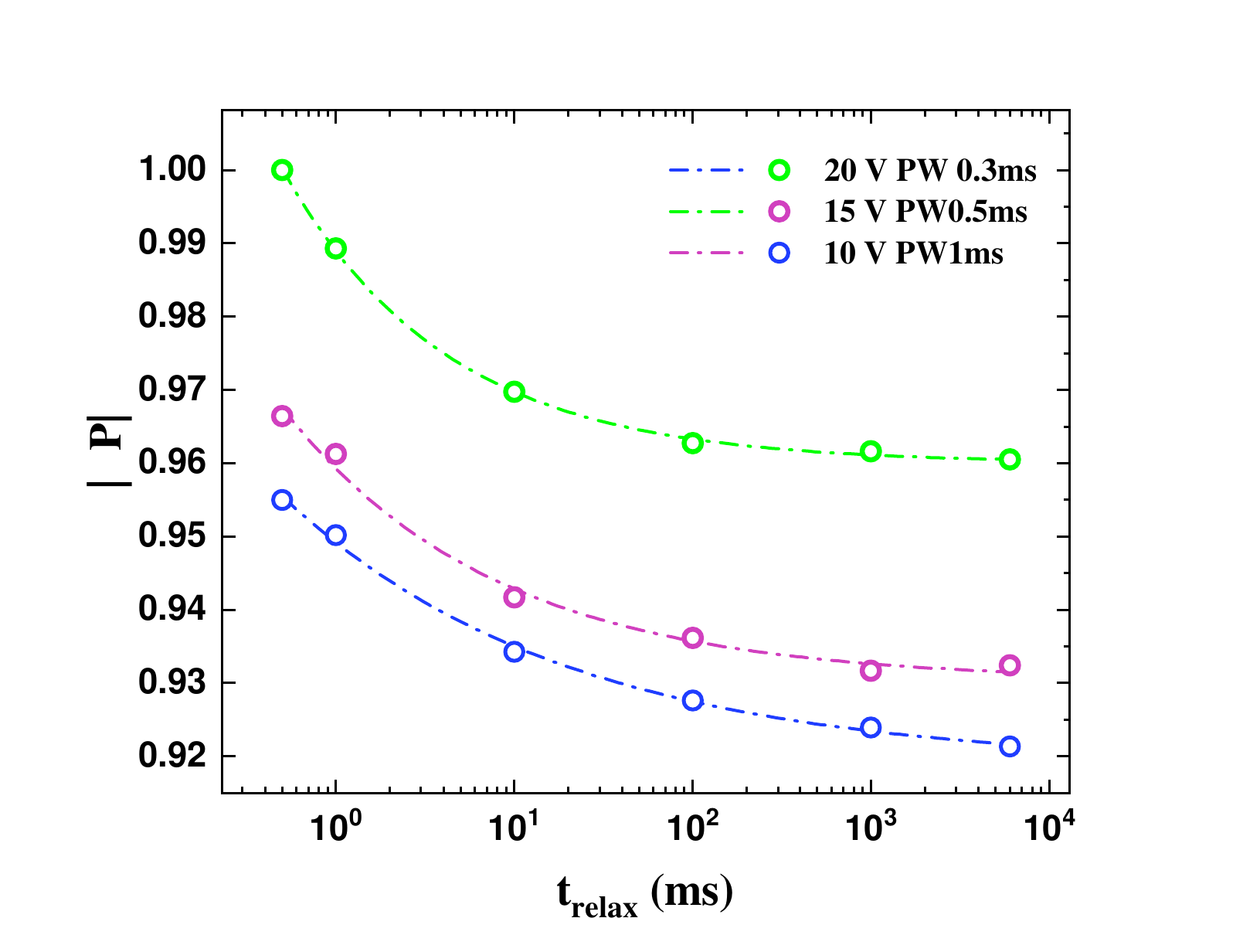}}
		\caption{Normalized switchable polarization $|\Delta P|$ as a function of the delay time $t_{\mathrm{relax}}$ between the write and read pulses, for different pulse conditions.}
	\end{figure*}
	
	At 6 V, a sharp Raman feature emerges near 196 cm$^{-1}$. This mode closely resembles the characteristic A1(LO) vibration reported for the tetragonal phase of PZT, indicating that the electric field promotes a significant tetragonal-like structural component. In addition, a Raman mode emerges near 270 cm$^{-1}$, assigned to an E + B$_1$-type mode that is commonly linked to the tetragonal phase \cite{deluca-2011}. Furthermore, we observed the intensity transfer within the E(TO2') band from $\sim$245 to $\sim$235 cm$^{-1}$ under bias that evidences a field-induced redistribution of local vibrational environments, consistent with phase-fraction rebalancing along the Zr/Ti compositional gradient in the MPB regime.
	
	Although the tetragonal character increases at high electric fields, the asymmetric spectral feature near 210 cm$^{-1}$ remains, showing that monoclinic phase still contribute to the Raman signal. Therefore, the applied field does not produce a full structural conversion but gradually shifts the phase equilibrium toward a tetragonal-like state while preserving a finite monoclinic fraction. Consequently, the 6 V spectrum is dominated by tetragonal-like vibrational signatures, showing an electric-field-driven evolution of Zr-rich MPB PZT from its rhombohedral [polarized along (111)] toward a tetragonal-like [polarized along (001)] configuration. Therefore, the raman spectroscopy indicates that the domain switching takes place through a state which is different from the initial and final state and different states can be accessed by varying the bias voltage parameters.

	\subsection{Polarization relaxation}
	
	The existence of a different intermediate state observed in Raman spectroscopy indicates that if polarization switching takes place through different switching pathways, the intermediate multidomain configurations generated during switching are expected to possess different metastabilities and therefore relax through different steps towards equilibrium. The impact of this can be seen on the relaxation dynamics of the system, especially on the relaxation time, domain configuration, stability, and metastable states \cite{liu2021atomistic,kim2025polarization}.
	
	We have studied the relaxation of switchable polarization by using a modified PUND pulse technique with fixed pulse width and increasing delay time from 0.1 ms to 10 s (represented as t$_{\mathrm{relax}}$) between the write and read pulse on (111) oriented PZT thin film. The difference between the P$_{\mathrm{sw}}$ and P$_{\mathrm{nsw}}$ gives the $\Delta$P, which represents the switchable polarization. The data has been fitted by the power law $|\Delta P| = P_{\mathrm{A}} + B\, t^{-n}$ where P$_{\mathrm{A}}$ is the polarization asymptote which accounts for the fraction of the switched polarization that is stabilized against back-switching by the pinning of domain walls at defect sites \cite{ganpule-2001}and $n$ is the decay exponent. The origin of power law dependence is from the lowering of nucleation energy due to the internal depolarizing field\cite{jo2006polarization,kim2005polarization}arising from the poor screening at the ferroelectric/electrode interface. 
	
	Figure 9 shows the normalized switchable polarization as a function of delay time. The pulse width has been chosen in such a way that it is close to the switching time scale at the corresponding field so that the polarization reaches the switching state. From the figure, we can see a systematic evolution of the extend of relaxation with voltage bias amplitude and pulse width. The exponent varies between 0.27 to 0.48 for the voltages 10-20 V the indicating that on applying higher voltage of shorter pulse width results in faster relaxation whereas for lower voltage applied for longer duration results in slower relaxation. We have also studied the variation of relaxation by varying the bias voltage amplitude, from 10 to 20 V, keeping the pulse width constant (Fig. S3 of supplemental materials document\cite{supplementarydoc}). The exponent is calculated to be between 0.26-0.29 pointing that the relaxation is following the same pathway. This indicates that both pulse width and amplitude combined together determines the pathway of relaxation to be followed and the relaxation kinetics depends on the switching conditions, with each state having different stability. 
	\begin{figure*}[!]
		\centering
		{\includegraphics[width=0.8\textwidth]{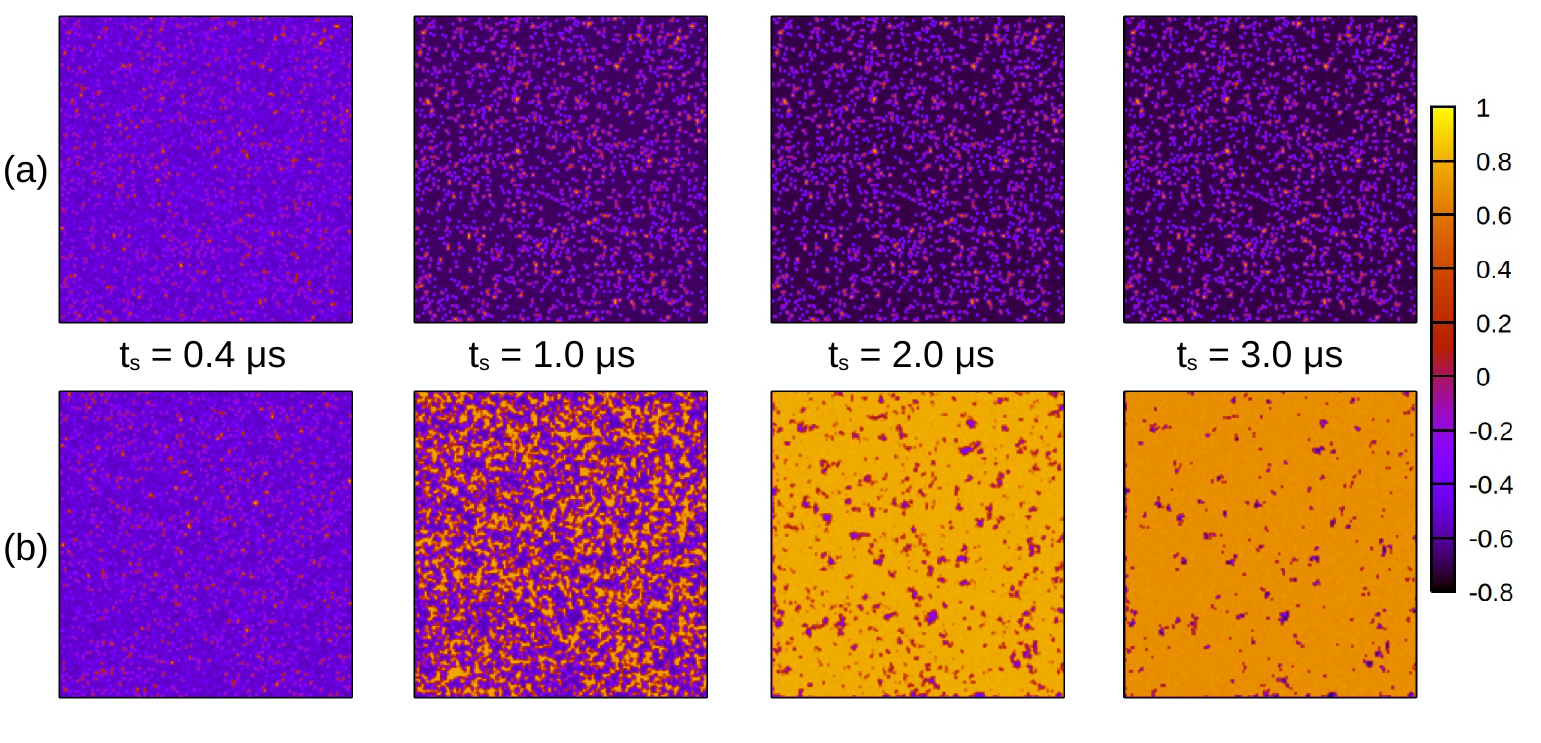}}
		\caption{Evolution of domain structure during relaxation for V$_{\mathrm{S}}$ = 10 V (a) switched OFF after 0.4 $\mu$s (t$_{\mathrm{off}}$=0.4 $\mu$s) and (b)switched OFF after 1 $\mu$s (t$_{\mathrm{off}}$=2 $\mu$s). }
	\end{figure*}
	
	\begin{figure*}[!]
		\centering
		{\includegraphics[width=0.8\textwidth]{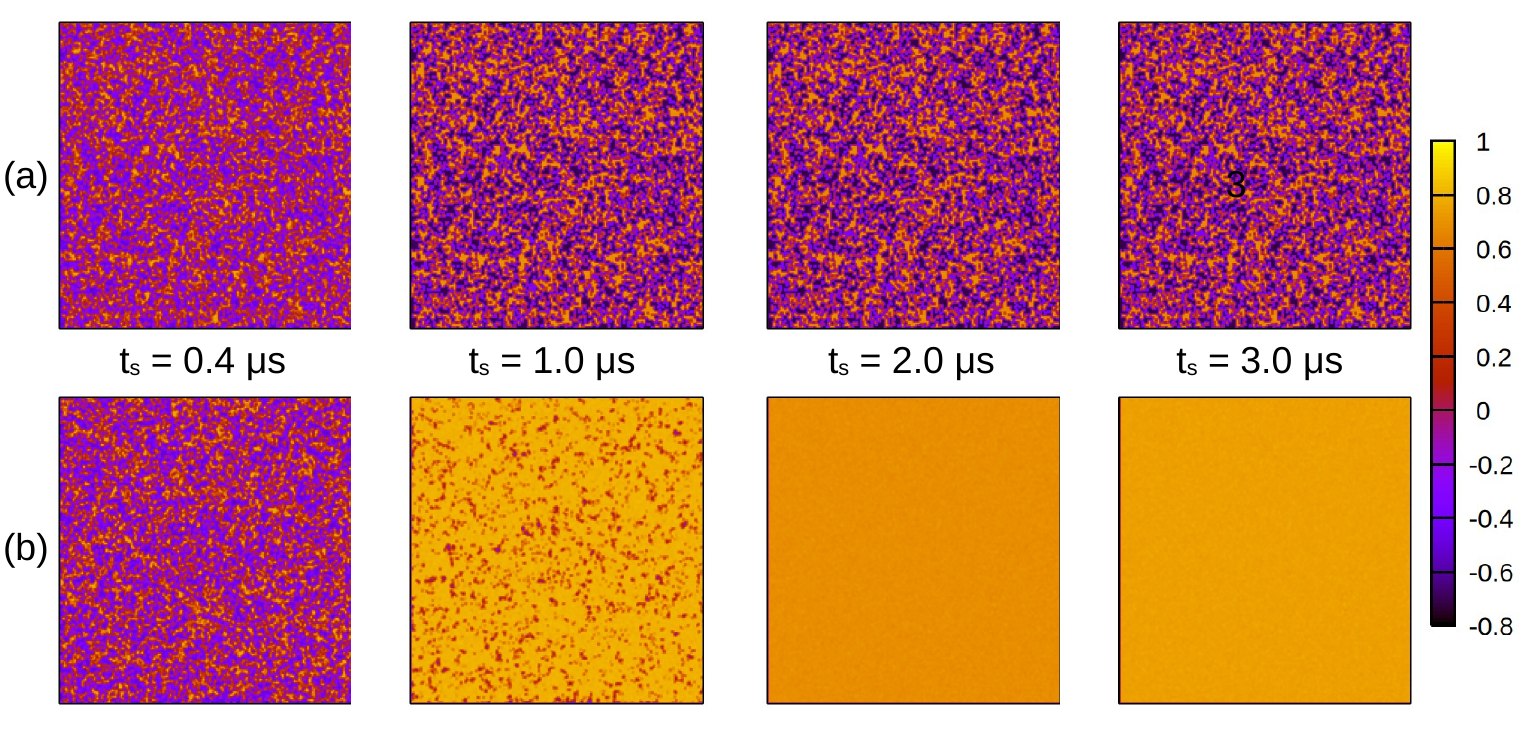}}
		\caption{ Evolution of domain structure during relaxation for V$_{\mathrm{S}}$ = 20 V (a) switched OFF after 0.4 $\mu$s (t$_{\mathrm{off}}$=0.4 $\mu$s) and (b)switched OFF after 1 $\mu$s (t$_{\mathrm{off}}$=1 $\mu$s).}
	\end{figure*}
	
	\begin{figure*}[!]
		\centering  
		\begin{subfigure}[b]{0.45\textwidth}\includegraphics[width=\textwidth]{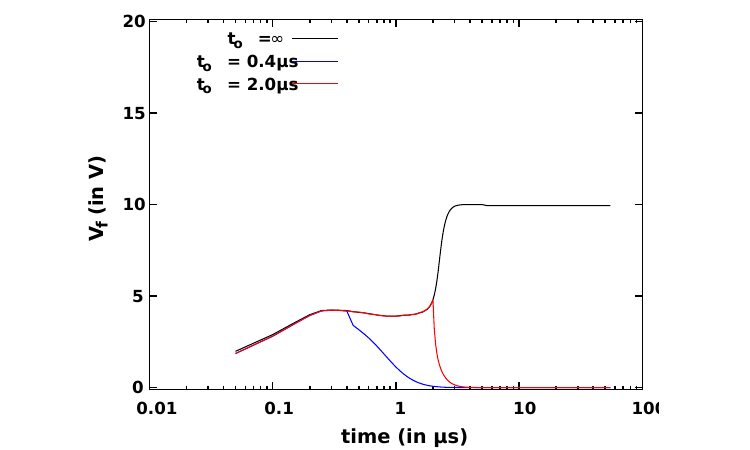}\caption{V$_{\mathrm{S}}$ = 10 V}\end{subfigure}
		\begin{subfigure}[b]{0.45\textwidth}\includegraphics[width=\textwidth]{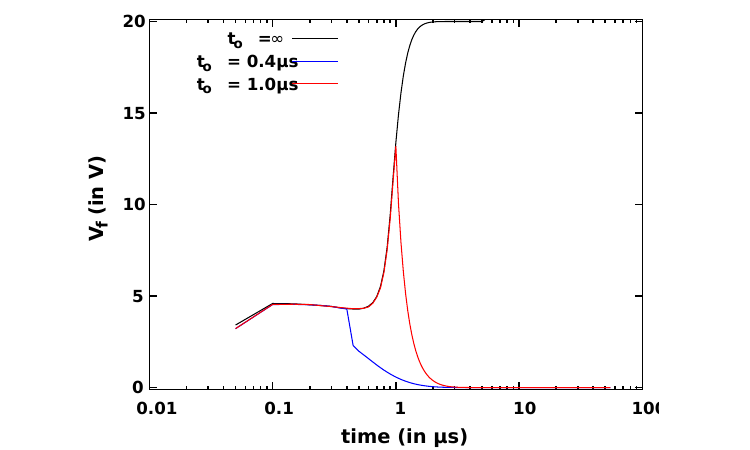}\caption{V$_{\mathrm{S}}$ = 20 V}\end{subfigure}
		\begin{subfigure}[b]{0.45\textwidth}\includegraphics[width=\textwidth]{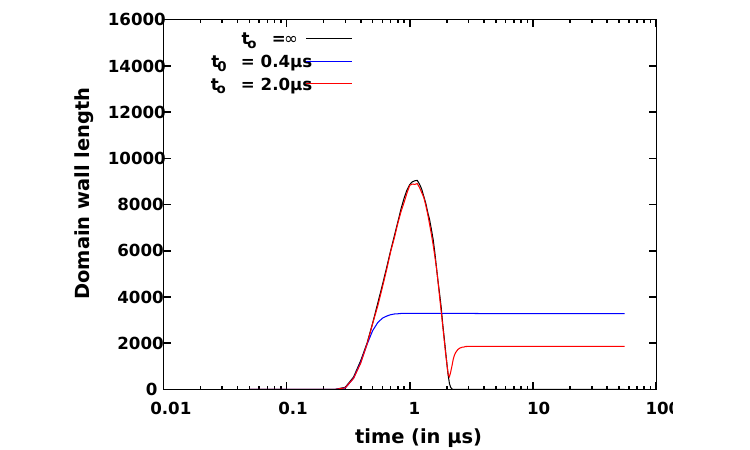}\caption{V$_{\mathrm{S}}$ = 10 V}\end{subfigure}
		\begin{subfigure}[b]{0.45\textwidth}\includegraphics[width=\textwidth]{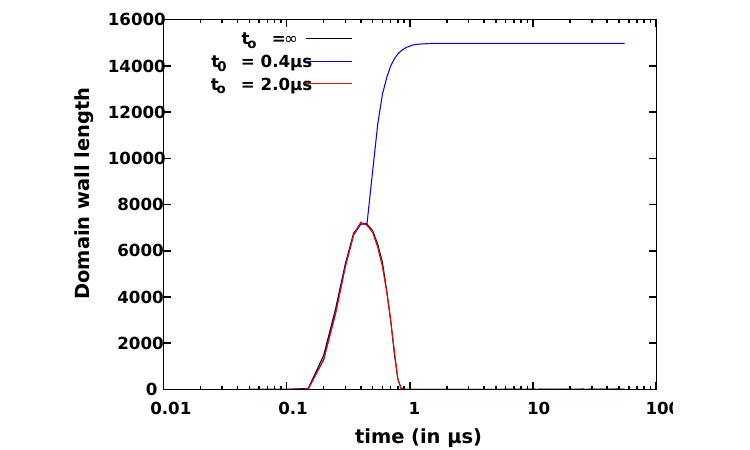}\caption{V$_{\mathrm{S}}$ = 20 V}\end{subfigure}
		\caption{(a) Evolution of V$_{\mathrm{f}}$ as the function of time during relaxation  for (a) V$_{\mathrm{S}}$ = 10 V and (b) V$_{\mathrm{S}}$ = 10 V. (b) Evolution of domain wall density during relaxation  for (a) V$_{\mathrm{S}}$ = 10 V and (b) V$_{\mathrm{S}}$ = 10 V. where $(t_{\mathrm{off}}=\infty)$ represents continuous application of (V$_{\mathrm{S}}$)}
	\end{figure*}
	
	The microscopic origin of the polarization relaxation and associated different pathways has been studied by using the time dependent Ginzburg-Landau equations of polarization dynamics using the model described in section III. In the simulation, the bias voltage was held ON until the system entered the transient negative capacitance state, identified by the region in which the V$_{\mathrm{F}}$ decreases, resulting in relaxation of the intermediate state. The bias voltage is then turned OFF and the average polarization is recorded as a function of time scale such that the we could capture the relaxation of the intermediate state of switching.
	
	The phase-field simulations provide further insight into the bias-voltage-dependent relaxation of the polarization after removal of the applied voltage. Figures 10 and 11 show the temporal evolution of the domain structure for \(V_{\mathrm{S}}=10\) and 20 V, respectively, for two different switching-off times. The simulations reveal that the domain configuration attained during the voltage pulse strongly influences the subsequent relaxation dynamics. This effect becomes particularly pronounced when the higher bias voltage is maintained for a longer duration.
	
	For \(V_{\mathrm{S}}=10\) V (Fig. 10), when the voltage is switched off at \(t_{\mathrm{off}}=0.4~\mu\mathrm{s}\), the polarization remains predominantly in a mixed-domain state throughout the investigated relaxation period. Although the morphology evolves after removal of the bias, extensive coexistence of oppositely polarized regions persists even at \(t=3~\mu\mathrm{s}\). Correspondingly, the domain-wall density remains appreciable. In contrast, when the 10 V pulse is maintained until \(t_{\mathrm{off}}=2.0~\mu\mathrm{s}\), the system has progressed substantially further along the switching pathway before removal of the field. Following removal of the voltage, the positively polarized regions grow at the expense of the oppositely polarized regions, resulting in progressive annihilation of domain walls. At longer relaxation times, only isolated regions of the reverse polarization remain.
	
	A stronger dependence on the pulse duration is observed for \(V_{\mathrm{S}}=20\) V (Fig. 11). For the shorter pulse (\(t_{\mathrm{off}}=0.4~\mu\mathrm{s}\)), a highly heterogeneous multidomain configuration is retained after the field is removed, and little net evolution toward a nearly single-domain state is evident over the displayed time interval. When the bias is instead maintained until \(t_{\mathrm{off}}=1.0~\mu\mathrm{s}\), rapid domain-wall motion and annihilation occur during relaxation. The system evolves from a densely intermixed domain configuration at \(0.4~\mu\mathrm{s}\) to an almost uniformly polarized state by approximately \(2~\mu\mathrm{s}\). Thus, the higher applied voltage, provided it acts for sufficient time, drives the system much closer to completion of polarization reversal and subsequently promotes rapid elimination of the remaining reverse domains.
	
	The connection between the electrical response and domain evolution is illustrated more quantitatively in Fig. 12. For \(V_{\mathrm{S}}=10\) V, switching the voltage off at \(t_{\mathrm{off}}=0.4~\mu\mathrm{s}\) leaves the system with a substantial domain-wall population, and the domain-wall length subsequently approaches a relatively high residual value. When the voltage is maintained until \(2.0~\mu\mathrm{s}\), the domain-wall length first reaches a pronounced maximum and subsequently decreases as neighboring domains coalesce and domain walls are annihilated. The peak in domain-wall length therefore identifies the stage at which the system contains the largest interfacial area between oppositely polarized regions.
	
	The behavior becomes even more distinct at \(V_{\mathrm{S}}=20\) V. Removal of the voltage at \(0.4~\mu\mathrm{s}\) arrests the system in a highly fragmented multidomain configuration, reflected by the large domain-wall length. In contrast, for \(t_{\mathrm{off}}=1.0~\mu\mathrm{s}\), the domain-wall length exhibits a sharp transient maximum followed by a rapid decrease toward nearly zero. This evolution is consistent with Fig. 11(b), where the initially dense domain pattern rapidly transforms into an almost uniformly polarized state. Importantly, the peak in domain-wall density occurs close to the rapid transient change in \(V_{\mathrm{F}}\), establishing a direct correlation between the electrical transient and the evolution of the domain structure.
	
	These results suggest that the bias-voltage dependence of the transient negative-capacitance response is governed not simply by the magnitude of polarization reversal, but by the non-equilibrium domain configuration present when the external bias is removed. The applied voltage and pulse duration determine the extent to which nucleation, domain growth, and domain-wall proliferation have developed before relaxation begins. When the bias is removed while a large population of mobile domain walls is present, subsequent domain growth and domain-wall annihilation can produce a rapid redistribution of polarization and hence a pronounced transient response. Conversely, removal of the bias either too early, before substantial domain evolution has occurred, or after the system has approached a nearly homogeneous switched state reduces the contribution associated with domain-wall-mediated relaxation. The simulations therefore support the interpretation that the observed bias-voltage-dependent transient negative capacitance originates from the voltage-dependent formation and subsequent relaxation of the labyrinthine-like domain structure.
	
	Thus, the Raman spectra, polarization relaxation experimental and theoretical simulation results indicate that the bias voltage parameters indeed determine the domain wall length at the intermediate states of switching, and by varying the bias voltage we can induce different states of switching. By tuning the voltage parameters, we can choose the pathway which would be of minimum energy barrier which yield maximum domain wall length. Since different intermediate states possess different domain-wall length as confirmed by phase-field modeling, and TNC scales with domain-wall length, the bias voltage dependent pathway selection provides a direct handle on the TNC magnitude. This establishes the point that the tunability of the TNC originates from the bias voltage dependence of the switching process itself and that the negative capacitance state can be engineered through external control of the domain wall area.
	
	\section{Summary and Conclusion}
	
	In summary, we have investigated the origin of the bias-voltage dependence of the transient negative capacitance in epitaxial Pb(Zr,Ti)O$_3$ ferroelectric capacitor having labyrinthine domain structure. Cross-sectional TEM, FESEM–EDS line scans, and XPS establish that the film exhibit a local variation in composition and lattice defects within its volume. Vector piezoresponse force microscopy shows that this compositional heterogeneity, together with strain, stabilizes a labyrinthine domain topology outside the conventional ultrathin regime in which these kind of domain topologies are normally observed. Raman spectroscopy combined with experimental and theoretical results of polarization relaxation indicates that switching can take place through different intermediate states. The bias voltage parameters strongly determines the domain wall length at the intermediate state of switching; since the domain wall area governs the magnitude of the transient negative capacitance the tunability of the TNC originates in the bias voltage dependence of the switching process itself. Therefore, the tunability of domain wall area by external parameters gives a possibility of direct engineering of the NC state of a multidomain ferroelectric capacitor.\\
	
\begin{center}
\textbf{ACKNOWLEDGMENTS}
\end{center}
	
	The support from the ANRF, Government of India, under the project SPR/2021/000131 is acknowledged. One of the authors (G.S.K.) acknowledges research fellowship under the DST-INSPIRE program of the Government of India. The author D.M. acknowledges financial support from the Anusandhan National Research Foundation (ANRF), Government of India (Grant No. ARNF/ARG/2025/007161/PS) and the India–Russia Joint Research Call, DST, Government of India (Grant No. DST/IC/RSF/2024/542). The authors also acknowledge helpful discussion with Ashish Garg, Indian Institute of Technology, Kanpur, Rupoma Bhattacharya, Aix-Marseille University, Marseille, and Manjunath Balagopalan, International Iberian Nanotechnology Laboratory, Braga.

\end{document}